\documentclass[10pt,twocolumn,letterpaper]{article}

\usepackage{iccv}
\usepackage{times}
\usepackage{epsfig}
\usepackage{graphicx}
\usepackage{amsmath}
\usepackage{amssymb}

\iccvfinalcopy 

\def\iccvPaperID{16} 

\ificcvfinal\fi

\usepackage[table,x11names]{xcolor}
\definecolor{Klein_Blue}{rgb}{0.0, 0.129, 0.6}
\definecolor{mygray}{gray}{0.95}
\usepackage{hyperref}       
\hypersetup{
    colorlinks=true,
    linkcolor=magenta,
    urlcolor=magenta,
    citecolor=Klein_Blue    
    }

\usepackage[export]{adjustbox}
\usepackage{hhline}
\usepackage{caption,tabularx,booktabs}
\usepackage{algpseudocode}
\usepackage[ruled,lined,linesnumbered]{algorithm2e}
    \usepackage{svg}

\algnewcommand{\LineComment}[1]{\State {\color{blue}\(\triangleright\) #1}}
\usepackage{subcaption}
\usepackage[export]{adjustbox}
\usepackage{amsfonts}
\usepackage{multirow}
\definecolor{mygray}{gray}{0.95}
\usepackage{commath}
\DeclareMathOperator*{\argmax}{arg\,max}

\newcolumntype{a}{>{\columncolor{mygray}}c}
\begin{document}

\title{Don't FREAK Out: A Frequency-Inspired Approach to Detecting Backdoor Poisoned Samples in DNNs}

\author{Hasan Abed Al Kader Hammoud$^1$
\and
Adel Bibi$^2$
\and
Philip H.S. Torr$^2$
\and
Bernard Ghanem$^1$\vspace{0.1cm}
\and 
$^1$ King Abdullah University of Science and Technology (KAUST) \quad $^2$ University of Oxford
}

\maketitle
\ificcvfinal\thispagestyle{empty}\fi

\begin{abstract}

In this paper we investigate the frequency sensitivity of Deep Neural Networks (DNNs) when presented with clean samples versus poisoned samples. Our analysis shows significant disparities in frequency sensitivity between these two types of samples. Building on these findings, we propose FREAK, a frequency-based poisoned sample detection algorithm that is simple yet effective. Our experimental results demonstrate the efficacy of FREAK not only against frequency backdoor attacks but also against some spatial attacks. Our work is just the first step in leveraging these insights. We believe that our analysis and proposed defense mechanism will provide a foundation for future research and development of backdoor defenses.

\end{abstract}
\section{Introduction}\label{sec:introduction}



Deep Neural Networks (DNNs) have revolutionized machine learning leading to remarkable advances in various domains such as autonomous vehicles \cite{Grigorescu2020ASO}, medical imagery analysis \cite{Danaee2017ADL}, and fraud detection \cite{Zhang2021HOBAAN}. The increased deployment of DNNs in life-critical applications, such as autonomous driving and medical diagnosis, raised concerns particularly with the uncovered vulnerabilities in the form of adversarial attacks.

One extremely insidious form of adversarial attacks is known as backdoor attacks. Backdoor attacks inject malicious behaviour through compromising the training procedure \cite{Li2020BackdoorLA,Doan2021LIRALI}, where at inference time, the attacker introduces a special input pattern, known as a trigger, inducing a targeted prediction. 
The true danger of backdoor attacks lies in their ability to bypass the normal validation procedures that ensure the accuracy and reliability of DNNs \cite{Gu2019BadNetsEB}. A backdoored model can behave normally on clean inputs and evade detection while misclassifying inputs that contain the trigger leading to severe consequences in high-stakes applications such as action recognition in surveillance systems \cite{hammoud2023look}.

\begin{figure}
    \centering
    \includegraphics[width=0.9\linewidth]{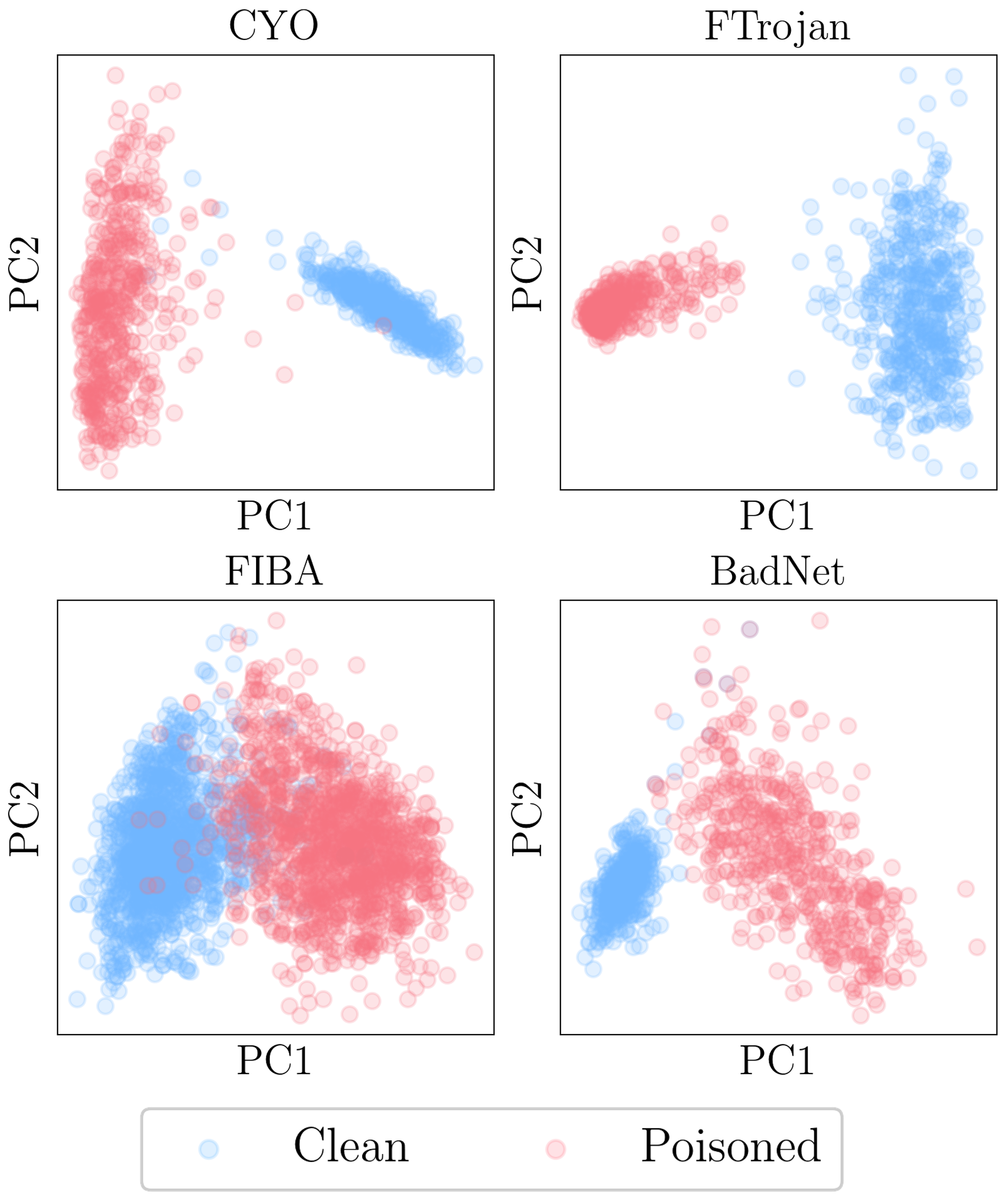}
    \caption{\textbf{FREAK PCA Features for Different Attacks.} The 2D PCA projection of the features extracted by FREAK are linearly separable which allows for the successful detection and separation of poisoned and clean samples. This observation holds true for frequency backdoor attacks (CYO, FTrojan and FIBA) and spatial backdoor attacks (BadNet). }
    \label{fig:my_label}
\end{figure}

Backdoor triggers were typically created in either the spatial \cite{Chen2017TargetedBA,Nguyen2021WaNetI,Doan2021LIRALI} or the latent domain \cite{Doan2021BackdoorAW,Yao2019LatentBA}. However, recent works have revealed that backdoor attacks could also be created in the frequency domain \cite{Hammoud2021CheckYO,Feng2021FIBAFB,Wang2021BackdoorAT}. Frequency-based backdoor attacks were shown to achieve high attack success rates with a capacity to elude state-of-the-art (SOTA) spatial and latent backdoor defenses. Given that adversaries have the ability to embed their poison in any frequency location across the input image channels, basic filtering techniques such as low-pass, band-pass, or high-pass filtering may not be able to eradicate the trigger.

In response to this challenge, researchers behind frequency-based backdoor attacks have proposed more advanced defenses. For instance, leveraging an autoencoder or JPEG compression \cite{Hammoud2021CheckYO} to manipulate the Fourier transform of tainted images and filter out the backdoor trigger was shown to be effective. On a different note, FTrojan \cite{Wang2021BackdoorAT} introduced two adaptive defenses that rely on either anomaly detection or signal smoothing. Nevertheless, these defenses are hampered by one or more limitations: (1) they function in the spatial domain (autoencoder \cite{Hammoud2021CheckYO}); (2) they can be circumvented by data augmentation (autoencoder and JPEG compression (\cite{Hammoud2021CheckYO}) and signal smoothing (\cite{Wang2021BackdoorAT}); (3) they cause significant drops in model accuracy on clean data (signal smoothing (\cite{Wang2021BackdoorAT}); or (4) they fail to detect the backdoor in the first place (anomaly detection \cite{Wang2021BackdoorAT}).

In this work, we analyze the distribution of the most sensitive frequency components when the DNN is presented with clean versus poisoned samples. Our analysis reveals that the frequency sensitivity to poisoned samples is considerably distinct from that of clean samples. Drawing on these findings, we present FREAK, a simple yet effective algorithm for identifying poisoned samples based on the distribution of the sensitive frequency components. Our algorithm achieves a high success rate in detecting poisoned samples 
while maintaining a low false positive rate. Surprisingly, FREAK is not only effective against frequency-based backdoor attacks but also against some spatial backdoor attacks. 


\section{Related Work}\label{sec:relatedwork}

In recent years, a variety of backdoor attacks have been proposed, each of which can differ in two key aspects: the method used to generate the trigger and whether or not the labels are manipulated. In response to these attacks, a number of backdoor defenses have been developed, which can be categorized as follows: (1) defenses aimed at detecting whether a model or a set of samples have been poisoned; (2) defenses aimed at mitigating the backdoor attack; and (3) defenses that aim to both detect and mitigate the attack simultaneously.

\textbf{Backdoor Attacks.} In early backdoor attack methods, backdoor triggers were designed in the spatial domain. For instance, \cite{Gu2019BadNetsEB} proposed poisoning the data by adding a black square in the corner of a few training samples. \cite{Liu2018TrojaningAO} solved an optimization problem to find an optimal backdoor trigger for a given mask, such as the square trigger introduced in \cite{Gu2019BadNetsEB}. However, as research progressed, the importance of invisible triggers that can bypass human inspection became evident, leading to the development of invisible backdoor attacks. This area of research has since evolved, with works such as \cite{Liu2020ReflectionBA,Chen2017TargetedBA,LSB,Chen2021UsePN,Wang2019NeuralCI,Doan2021LIRALI,Bagdasaryan2021BlindBI,Zhang2021PoisonIR,Ren2021SimtrojanSB,Liao2020BackdoorEI} paving the way. \cite{Chen2017TargetedBA} proposed blending the backdoor trigger with clean images instead of stamping it. \cite{LSB} and \cite{Li2021InvisibleBA} adopted least significant bit and textual string encoding algorithms from steganography to poison the data, respectively. \cite{Nguyen2021WaNetI} used image warping as a poisoning technique, while \cite{Doan2021LIRALI} emphasized the importance of having learnable transformations to embed an optimal backdoor trigger into the poisoned samples. \cite{Chen2021UsePN} showed that procedural noise, such as Gabor and Perlin noise, could be used as a backdoor trigger. \cite{Turner2018CleanLabelBA,Barni2019ANB,Turner2019LabelConsistentBA} suggested clean-label backdoor attacks, which apply a backdoor trigger without manipulating the class label of the images.

More recent works suggested exploring alternative domains. For instance, \cite{Doan2021BackdoorAW} generated imperceptible backdoor triggers by minimizing the Wasserstein distance (\cite{Kolouri2019GeneralizedSW}) between latent representations of clean and poisoned samples. \cite{Zeng2021RethinkingTB} analyzed the characteristics of spatial backdoor attacks in the Fourier domain and present a technique to generate smooth spatially visible triggers that are smooth in the frequency domain. Finally, and most relevant to our work, \cite{Hammoud2021CheckYO,Feng2021FIBAFB,Wang2021BackdoorAT} showed the power of embedding backdoor attacks in the frequency domain. \cite{Hammoud2021CheckYO} utilized the concept of Fourier heatmaps from \cite{Yin2019AFP} to detect the DNN’s most sensitive frequency bases, which are then used to mount the poisoning information. \cite{Feng2021FIBAFB} suggested blending the low-frequency content of a trigger image with those of clean samples to generate poisoned data. \cite{Wang2021BackdoorAT} converted the color channels from RGB to YUV representation, after which a mix of mid- and high-frequency components is poisoned to bypass possible low-pass or high-pass filtering.

\textit{In this work we analyze the properties of frequency-based backdoor attacks. Based on the uncovered properties we propose a new defense.}

\textbf{Backdoor Defenses:} As mentioned above, backdoor defenses try to detect the attack \cite{Gao2019STRIPAD,Liu2019ABSSN,Huster2021TOPBD,Zheng2021TopologicalDO,Hammoudeh2021SimpleAD,Chen2021DePoisAA,Soremekun2020ExposingBI,Tang2021DemonIT,Jin2020AUF}, mitigate the attack \cite{Liu2018FinePruningDA,Liu2017NeuralT,Cheng2020DefendingAB,Li2020RethinkingTT}, or both detect and mitigate the attack \cite{Tran2018SpectralSI,Chen2019DetectingBA,Wang2019NeuralCI,Guo2019TABORAH,Liu2019ABSSN,Doan2020FebruusIP,Hayase2021SPECTREDA,Jiang2021InterpretabilityGuidedDA,Qiao2019DefendingNB,Chen2019DeepInspectAB}.

Early backdoor defenses, such as neural cleanse \cite{Wang2019NeuralCI}, observed that backdoor attacks create an anomalously small distance between all classes and the poisoned class. On the basis of this observation, the authors proposed solving an optimization problem to detect whether a model has been poisoned after which the backdoor trigger is reverse engineered. Later, improved versions of this defense were introduced by TABOR \cite{Guo2019TABORAH} and ABS \cite{Liu2019ABSSN}.

Other backdoor attacks focused on understanding the activations of backdoor attacked models. Fine pruning \cite{Liu2018FinePruningDA} argued that backdoor attacks could be detected by pruning neurons that are dormant in the presence of clean samples; activation clustering \cite{Chen2019DetectingBA} and \cite{Tran2018SpectralSI, Hayase2021SPECTREDA} applied cluster analysis and robust statistics to detect and mitigate backdoor attacks. \cite{Doan2020FebruusIP} observed that backdoor attacks shift the network's attention away from the object, and therefore proposed applying image restoration to reconstruct the spatially poisoned region. Recently, \cite{Zheng2021TopologicalDO} used homology from topological analysis to uncover structural abnormalities unique to poisoned models.

Unfortunately, existing defenses against frequency backdoor attacks are very scarce and have been shown to fail in certain scenarios. For example, \cite{Hammoud2021CheckYO} proposed using an autoencoder or JPEG compression to manipulate the Fourier transform of poisoned images and hence neutralize the effect of the backdoor trigger. \cite{Wang2021BackdoorAT} proposed applying preprocessing techniques such as Gaussian and Wiener filtering to remove the backdoor. However, these defenses cause a huge drop in clean data accuracy or fail to neutralize the backdoor attack.

\textit{Considering the limited success of existing defenses in defending against both spatial and frequency backdoor attacks and the critical importance of detecting poisoned samples, FREAK stands out as an effective and necessary addition to the arsenal of defenses against backdoor attacks.}

\section{Properties of Frequency Backdoor Attacks}\label{sec:introduction}

\begin{figure*}[t!]
    \centering
    \includegraphics[width=0.9\textwidth]{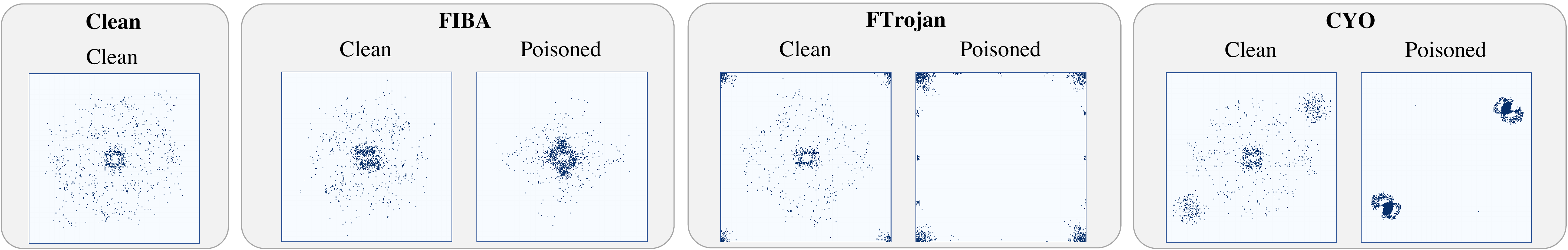}
    \caption{\textbf{Visualizing the Indices of the top-k Most Sensitive Frequencies.} By visualizing the top-$1000$ indices of the FREAK gradient, $\nabla_{\text{FREAK}}(x)$, we can identify the frequency bases that a neural network is most sensitive to for a particular input. We show these indices for a Clean model with a clean input, and for models that have been poisoned with FIBA, FTrojan, and CYO attacks, with both clean and poisoned inputs. This allows us to gain insight into the specific frequencies that are most important to each model and how different attacks affect the network's sensitivity.}
    \label{fig:analysis1}
\end{figure*}

\begin{figure}[t!]
    \centering
    \includegraphics[width=\linewidth]{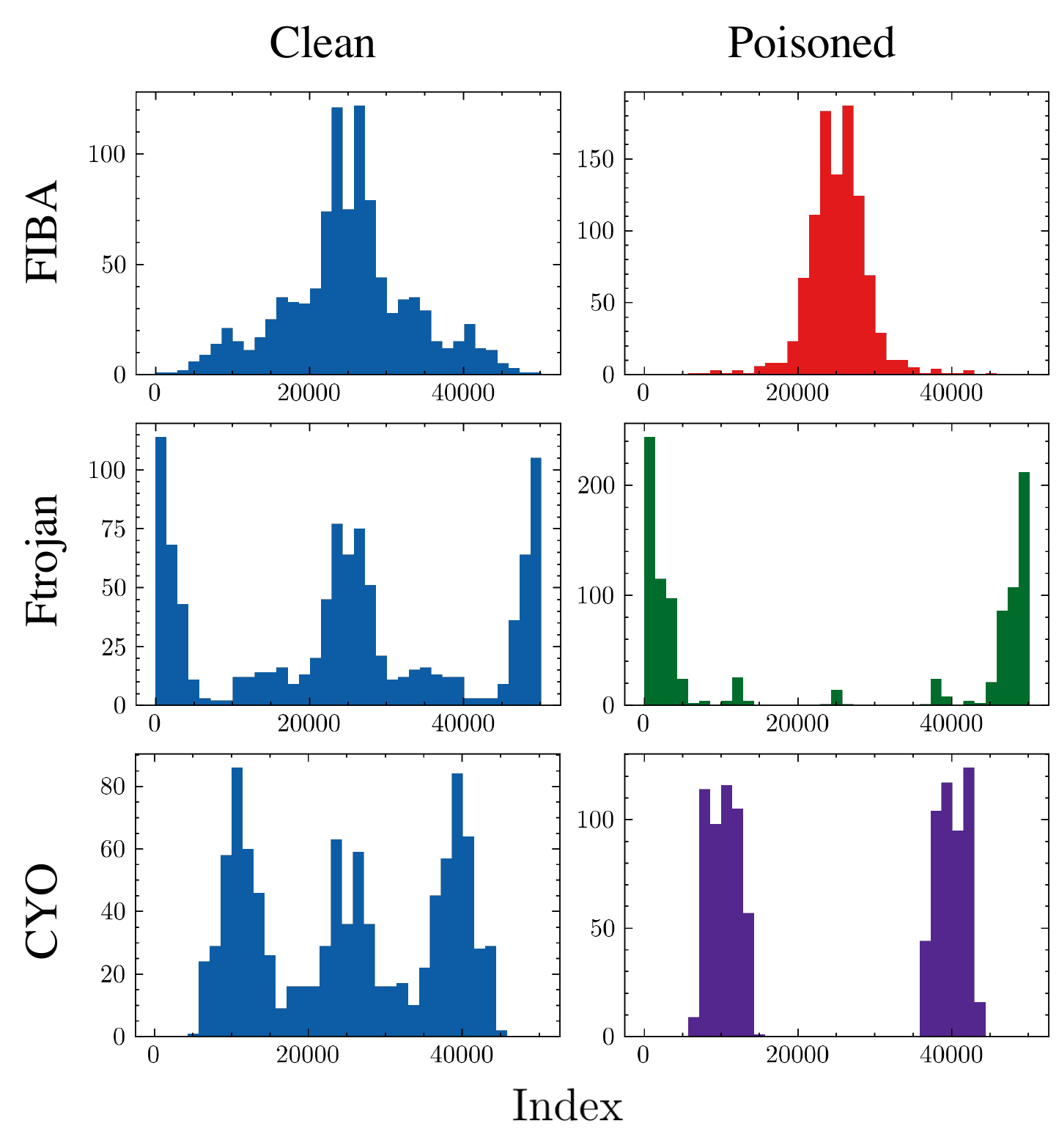}
    \caption{\textbf{Visualizing the Distribution of the Indices of the top-k Most Sensitive Frequencies.} Analyzing the distribution of the top-k most sensitive frequency indices can help us detect whether a sample has been poisoned. We compare the distribution shifts for clean and poisoned samples using three attacks: FIBA, FTrojan, and CYO. Backdoored models experience a drastic shift in frequency sensitivity in the presence of the backdoor trigger. This provides valuable insights into the effects of backdoor attacks on the network's sensitivity to frequency bases.}
    \label{fig:analysis2}
\end{figure}

\begin{figure*}[ht!]
    \centering
    \includegraphics[width=0.9\textwidth]{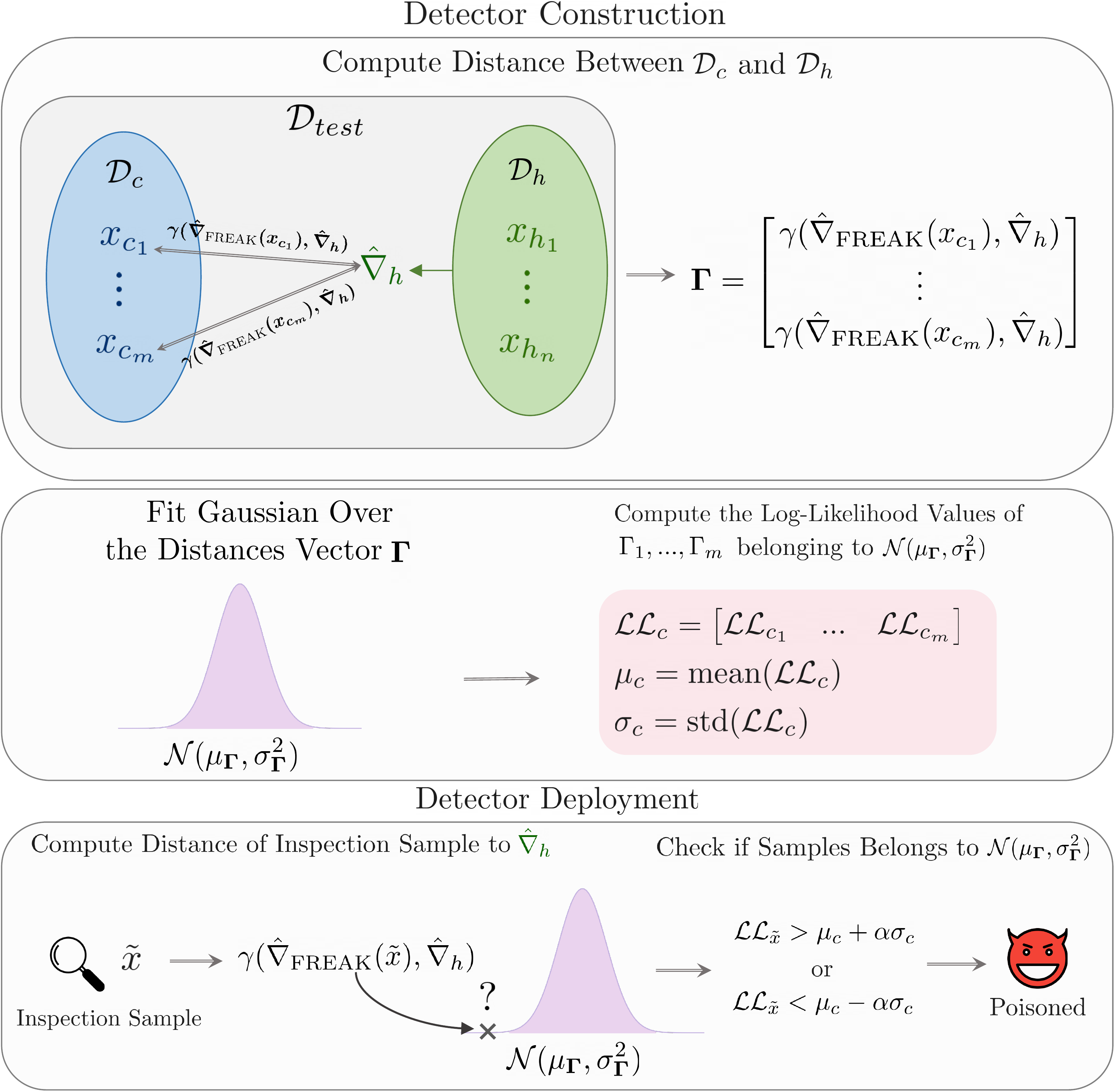}
    \caption{\textbf{FREAK Poisoned Sample Detection.} To construct FREAK detector, we first find the distance in Equation \ref{freak_eq} between samples from a clean experimental set $\mathcal{D}_c$ and the average distribution of the indices top-$k$ held-out set $\mathcal{D}_h$, referred to as $\hat{\nabla}_h$. The obtained distances are stored in a vector $\mathbf{\Gamma}$, whose rows represent the distance of one sample from $\mathcal{D}_c$ to $\hat{\nabla}_h$. Next, we fit a Gaussian distribution over the rows of $\mathbf{\Gamma}$, referred to as  $\mathcal{N}({\mu}_\Gamma, {\sigma}^2_{\Gamma})$, and compute the log-likelihood values of the rows of $\mathbf{\Gamma}$ belonging to that distribution. We store the mean and the standard deviation of the log-likelihood values in $\mu_c$ and $\sigma_c$. When a new sample $\tilde{x}$ is to be inspected, we compute the distance of $\tilde{x}$ to $\hat{\nabla}_h$ and compute the likelihood value of this distance belonging to $\mathcal{N}({\mu}_\Gamma, {\sigma}^2_{\Gamma})$, if the value falls within $\alpha$ standard-deviations of the mean then the sample is clean, otherwise it is poisoned.  } 
    \label{pipeline}
\end{figure*}

\paragraph{Analysis. }
Existing frequency-based backdoor attacks, \eg, FIBA \cite{Feng2021FIBAFB}, CYO \cite{Hammoud2021CheckYO}, and FTrojan \cite{Wang2021BackdoorAT}, embed poisoned information into specific frequency bases. Therefore, one would expect that in the presence of the backdoor trigger, such models would attend to the poisoned bases for classifying the poisoned sample. To verify this, we devise an approach similar to the spatial attention technique, GradCAM \cite{Selvaraju2019GradCAMVE}. The idea is to compute the gradient of the maximal logit of the network with respect to the Fourier transform, specifically, the Fourier magnitude. 
Afterwards, we visualize the indices of the $k$ most sensitive frequency bases, \ie, top-$k$ gradient values.

Formally, let $f_\theta : \mathcal{X} \rightarrow \mathcal{Y}$ be a classifier parameterized by $\theta$ mapping images $x\in \mathcal{X}$ to class labels $y\in \mathcal{Y}$. We denote the most probable class prediction for image $x$ by $c_A$, where $c_A = \argmax \limits_{c \in \mathcal{Y}} f_\theta^c(x)$. Let $\mathcal{G}_\eta: \mathcal{X}\rightarrow \mathcal{X}$ denote an attacker-specific poisoned image generator which is parameterized by $\eta$. Finally, let $\mathcal{F}(x)$ be the 2D Discrete Fourier Transforms (DFT) of an image $x$. 
The gradient we are interested in computing is

\begin{equation}
    \nabla_{\text{FREAK}}(x) = \nabla_{\mid\mathcal{F}(x)\mid}~f_\theta^{c_A}(x).
\end{equation}
The above quantity expresses the sensitivity of the classifier's prediction with respect to the Fourier magnitude. 

We are interested in comparing the above gradient for clean samples, $x_c$ and their poisoned counterparts $x_p = \mathcal{G}_\eta(x_c)$. Figure \ref{fig:analysis1}, presents a binary map that highlights the indices of the top-$k$ values of $\nabla_{\text{FREAK}}(x_c)$ and $ \nabla_{\text{FREAK}}(x_p)$ where as Figure \ref{fig:analysis2} 
shows the distribution of those indices. 
We make the following key observation; \emph{the frequency bases the network attends to for predicting clean samples differ drastically from those for poisoned samples.} This observation is the fundamental observation behind FREAK.

\paragraph{FREAK Defense. } 
 During inference time, the victim is presented with a sample which may or may not be poisoned. Since the victim has access to clean samples from their test set, we find that a simple mechanism to detect poisoned samples is computing a statistical metric, such as Wasserstein distance, between the distribution of the indices of the top-$k$ most sensitive frequency bases of a sample under inspection sample and that of clean samples. More precisely, we define $\hat{\nabla}_{\text{FREAK}}(x)$ as 


\begin{equation}
\begin{aligned}
\small
\hat{\nabla} _{\text{FREAK}}(x)[i,j] = \begin{cases}
                          1 & \text{if } {\nabla} _{\text{FREAK}}(x)[i,j] \in \text{top-}k({\nabla} _{\text{FREAK}}(x)), \\
                          0 & \text{otherwise}.
                        \end{cases}
\end{aligned}
\end{equation}

\noindent \ie $\hat{\nabla}_{\text{FREAK}}(x)$ is a binary matrix with value 1 in the locations of the top-$k$ values of $\nabla_{\text{FREAK}}(x)$. The distance between the distribution of the indices of the top-$k$ most sensitive frequency bases can be written as, 
\begin{equation}\label{freak_eq}
    \gamma(x,y) = d(\text{pool}(x),\text{pool}(y)),
\end{equation}
where $d$ is the Wasserstein distance and $\text{pool}$ is a simple sum-pooling function 
that aggregates values to obtain a distribution like mapping out of the binary matrices. 



\vspace{0.3cm}
\noindent The recipe for FREAK is visually presented in Figure \ref{pipeline} and is described below.
\begin{enumerate}
    \item From the test set $\mathcal{D}_{test}$, create two subsets of samples, a held-out set $\mathcal{D}_h = \{x_{h_{1}},x_{h_{2}},..,x_{h_{n}}\}$ and a clean-experimental set $\mathcal{D}_c = \{x_{c_1},x_{c_2},..,x_{c_m}\}$  where $\mathcal{D}_c\cap \mathcal{D}_h = \phi$ and $\mathcal{D}_c \cup \mathcal{D}_h \subseteq \mathcal{D}_{test}$.
    \item Compute   $\hat{\nabla}_{h} = \frac{1}{n} \sum_{j=1}^{n}
    \hat{\nabla}_{\text{FREAK}} (x_{h_j})$. 
    \item Compute and store the distance vector $\mathbf{\Gamma}_{i} = \gamma(\hat{\nabla}_{\text{FREAK}}(x_{c_i}),\hat{\nabla}_h)$ for $ i=1,..,m$. 
    \item Fit a Gaussian distribution over the rows of $\mathbf{\Gamma}$. The obtained distribution is denoted by $\mathcal{N}(\mu_\mathbf{\Gamma},\sigma_\mathbf{\Gamma}^2)$ where ${\mu}_\Gamma$ and ${\sigma}_\Gamma^2$ are the mean and covariance of the fit Gaussian distribution.
    \item Compute the log-likelihood values of the samples of $\mathcal{D}_c$ belonging to the previously fit Gaussian, 
\begin{equation}
\small
\begin{aligned}
\mathcal{L}\mathcal{L}_{c_i} = 
\text{log } p(x \mid \mu_\mathbf{\Gamma}, \sigma_\mathbf{\Gamma}^2) = -\frac{1}{2} \text{log}(2\pi\sigma_\mathbf{\Gamma}^2) - \frac{(x - \mu_\mathbf{\Gamma})^2}{2\sigma_\mathbf{\Gamma}^2}
\end{aligned}
\end{equation}

    for $i=1,...,m$, and store the mean $\mu_c$ 
    and the standard deviation $\sigma_c$, 
    of the log-likelihood scores.
    \item When a new sample $\Tilde{x}$ is presented for inspection, calculate the distance  $\gamma(\Tilde{x}, \hat{\nabla_h})$ and compute the log-likelihood of the vector belonging to the previously fit Gaussian distribution.  If $\mathcal{L}\mathcal{L}_{\Tilde{x}}>\mu_c+\alpha\sigma_c$ or $\mathcal{L}\mathcal{L}_{\Tilde{x}}<\mu_c-\alpha\sigma_c$ then the sample is poisoned, otherwise it's clean.
\end{enumerate}

\section{Experiments}\label{sec:experiments}

\begin{table*}[th!]
\centering
\renewcommand{\arraystretch}{1.1}
\scalebox{0.85}{
\begin{tabular}{@{}cccccccc@{}}
\hhline{--------} 
         \multicolumn{3}{c}{\textbf{FIBA}~\cite{Feng2021FIBAFB}}  & \multicolumn{3}{a}{\textbf{FTrojan}~\cite{Wang2021BackdoorAT} }      & \multicolumn{2}{c}{\textbf{CYO}~\cite{Hammoud2021CheckYO}}           \\ \hhline{--------} 
         $h$  & $\alpha$  & PSNR$\uparrow$/SSIM$\uparrow$  & {Locations} & {Magnitude} & PSNR$\uparrow$/SSIM$\uparrow$ & $k$ & \multicolumn{1}{l}{PSNR$\uparrow$/SSIM$\uparrow$} \\ \hhline{--------} 

  50 &    0.2    &    23.98/0.9010       &     (223,224), (111,111)      &     30.0      &     44.89/0.9943      & 1000 &   49.51/0.9981                            \\ \bottomrule
\end{tabular}}
\vspace{5pt}
\caption{\textbf{Parameters of Frequency Backdoor Attacks.} The parameters of each frequency backdoor attack are chosen such that an ASR $>95\%$ is achieved. These parameters, along with the invisibility metrics for each attack, are summarized here.}
\label{hyperparameters}
\end{table*}

\subsection{Experimental Setup \& Metrics} 
\noindent \textbf{Setup.} Similar to \cite{Hammoud2021CheckYO, Doan2021LIRALI}, we conduct our experiments on ImageNet dataset \cite{Russakovsky2015ImageNetLS}. All models are trained using a ResNet18 trained from scratch using an SGD optimizer with initial learning rate of 0.1 that decays by a factor of 0.25 every 15 epochs. The poisoning rate is fixed to 5.0\%.

\noindent \textbf{Backdoor Attack Metrics.}  To evaluate the performance of the trained backdoor attacked models, we use two commonly used metrics: clean data accuracy (CDA), which measures the DNN’s performance on clean samples, and attack success rate (ASR), which measures the effectiveness of the backdoor attack in instigating the target label. A good backdoored model should have a high ASR and a high CDA.

\noindent \textbf{Detector Metrics.} To evalute the performance of the proposed FREAK detector, we use True Positive Rate (TPR) and False Positive Rate (FPR) as metrics. TPR is a measure of how often a detector correctly identifies a poisoned sample. It is calculated as the number of true positive instances divided by the total number of positive instances. FPR is a measure of how often a detector incorrectly identifies a clean samples as poisoned. It is calculated as the number of false positive results divided by the total number of negative instances. Both TPR and FPR are important metrics in evaluating the performance of a detector. TPR helps us to assess how effective the detector is at identifying poisoned samples, while FPR helps us to identify cases where the detector is misclassifying clean samples as poisoned. A good detector should have a high TPR and a low FPR.

\noindent \textbf{Invisibility Metrics.} Following \cite{Hammoud2021CheckYO}, we measure the imperceptibility of an attack using peak signal-to-noise ratio (PSNR) and structural similarity (SSIM). SSIM is a perceptual metric that compares the structural similarity between two images, while PSNR measures the peak signal-to-noise ratio between the two images. A higher SSIM or PSNR value indicates a higher quality image, \textit{i.e poisoned image looks close to clean one}, while a lower value indicates a more noticeable attack.

\subsection{FREAK against Frequency Backdoor Attacks}

We evaluate our backdoor defense against three frequency backdoor attacks, namely, CYO \cite{Hammoud2021CheckYO}, FTrojan \cite{Wang2021BackdoorAT}, and FIBA \cite{Feng2021FIBAFB}. As mentioned earlier, the models are trained from scratch using a poisoning rate of 10.0\%. Table \ref{hyperparameters} shows the invisibility metrics (PSNR and SSIM) and hyperparameters selected for each attack. The hyperparameters were chosen to achieve an ASR $>95.0\%$. 

To test our proposed defense, we fix $\alpha=1$, $|\mathcal{D}_h|= 32$, $|\mathcal{D}_c| = 128$, $\beta = 12$ (pooling filter size), and $k=5000$. We randomly select 5000 samples from $\mathcal{D}_{\text{test}}$ to be poisoned and another 5000 samples to compute the false positive rate. 

FREAK has demonstrated remarkable capabilities in achieving a remarkably high true positive rate (TPR) while simultaneously maintaining an impressively low false positive rate (FPR) in response to all frequency backdoor attacks. Specifically, against CYO \cite{Hammoud2021CheckYO} and FTrojan \cite{Wang2021BackdoorAT}, FREAK attains a TPR that is close to perfect, i.e., 100\%, with an accompanying FPR that is insignificantly close to zero. However, because FIBA \cite{Feng2021FIBAFB} corrupts low-frequency data, which typically coincides with the frequencies that a clean network processes, the TPR decreases to 90\% with an FPR of 5\%. Further details about the results obtained for ResNet34 and ablations of different hyperparameters can be found in the supplementary material.

\begin{table}[h!]
\centering
\scalebox{0.9}{
\begin{tabular}{@{}ccc@{}}
\cmidrule(l){2-3}
\textbf{}        & \textbf{TPR (\%)} & \textbf{FPR (\%)} \\ \midrule
\textbf{CYO}     & 99.25             & 1.56              \\
\textbf{FTrojan} & 100.00            & 1.39              \\
\textbf{FIBA}    & 90.15             & 5.31              \\ \bottomrule
\end{tabular}}
\caption{\textbf{FREAK Against Frequency Backdoor Attacks.} FREAK proves to be capable of achieving a high TPR while maintaining a low FPR against frequency backdoor-attacks.}
\end{table}

\subsection{FREAK against Spatial Backdoor Attacks}

We subjected FREAK to a series of spatial backdoor attacks, including BadNet \cite{Gu2019BadNetsEB}, Blend \cite{Chen2017TargetedBA}, SIG \cite{Barni2019ANB}, and WaNet \cite{Nguyen2021WaNetI}, and summarized the outcomes in Table \ref{spatial}. Interestingly, FREAK was able to achieve a high true positive rate against BadNet and SIG while maintaining a low false positive rate; however, this was not the case for Blend and WaNet, where a significant decline in performance was observed. We discuss this further in the limitations section. The robust performance of FREAK against SIG attacks may be due to the fact that the sinusoidal signal utilized for poisoning the model is of high frequency, creating distinct artifacts in the frequency domain compared to a clean sample.

\begin{table}[h!]
\centering
\scalebox{0.9}{
\begin{tabular}{@{}ccc@{}}
\cmidrule(l){2-3}
                & \textbf{TPR (\%)} & \textbf{FPR (\%)} \\ \midrule
\textbf{BadNet} & 96.60             & 2.73              \\
\textbf{SIG}    & 84.51             & 4.74              \\
\textbf{WaNet}  & 2.34              & 1.95              \\
\textbf{Blend}  & 9.11              & 3.91              \\ \bottomrule
\end{tabular}}
\caption{\textbf{FREAK Against Spatial Backdoor Attacks.} FREAK proves to be capable of achieving a high TPR while maintaining a low FPR on BadNet and SIG, however, this is not the case for WaNet and Blend.}
\label{spatial}
\end{table}

\section{Conclusion}\label{sec:conclusion}

\textbf{Limitations.} While our analysis sheds light on the behavior of neural networks when presented with clean and poisoned samples from a frequency-domain standpoint, we acknowledge that the proposed FREAK method is only one possible approach for leveraging these insights, and it may not necessarily be the most optimal. Furthermore, although we opted to use Wasserstein distance, other more suitable distances may be available for this particular problem. Lastly, although our findings encompass all frequency-based backdoor attacks, we acknowledge that we only evaluated a fraction of spatial attacks, and thus further research is required to obtain a more comprehensive assessment.

\textbf{Conclusion.} In conclusion, this paper presents a comprehensive investigation into the frequency sensitivity of Deep Neural Networks when exposed to clean versus poisoned samples. Our results reveal significant differences in frequency sensitivity between these two types of samples. Based on these findings, we propose FREAK, a novel frequency-based poisoned sample detection algorithm that is both simple and effective. Our experimental results demonstrate that FREAK is not only successful against frequency-based backdoor attacks but also some spatial attacks. While our work represents a critical first step towards leveraging these insights, we anticipate that our analysis and proposed defense mechanism will establish a basis for future research and development of backdoor defenses. One possible future direction is sample purification which is presented in the supplementary material.

\section{Acknowledgements} 

This work was supported by the King Abdullah University of Science and Technology (KAUST) Office of Sponsored Research through the Visual Computing Center (VCC) funding, the SDAIA-KAUST Center of Excellence in Data Science and Artificial Intelligence (SDAIA-KAUST AI), and UKRI grant: Turing AI Fellowship EP/W002981/1. We also thank the Royal Academy of Engineering and FiveAI for their support. Adel Bibi has received funding from the Amazon Research Awards.

{\small
\bibliographystyle{ieee_fullname}
\bibliography{egbib}
}

\clearpage
\appendix
\onecolumn
\section{Future Directions:}

\begin{figure}[h!]
    \centering
    \includegraphics[width=0.85\textwidth]{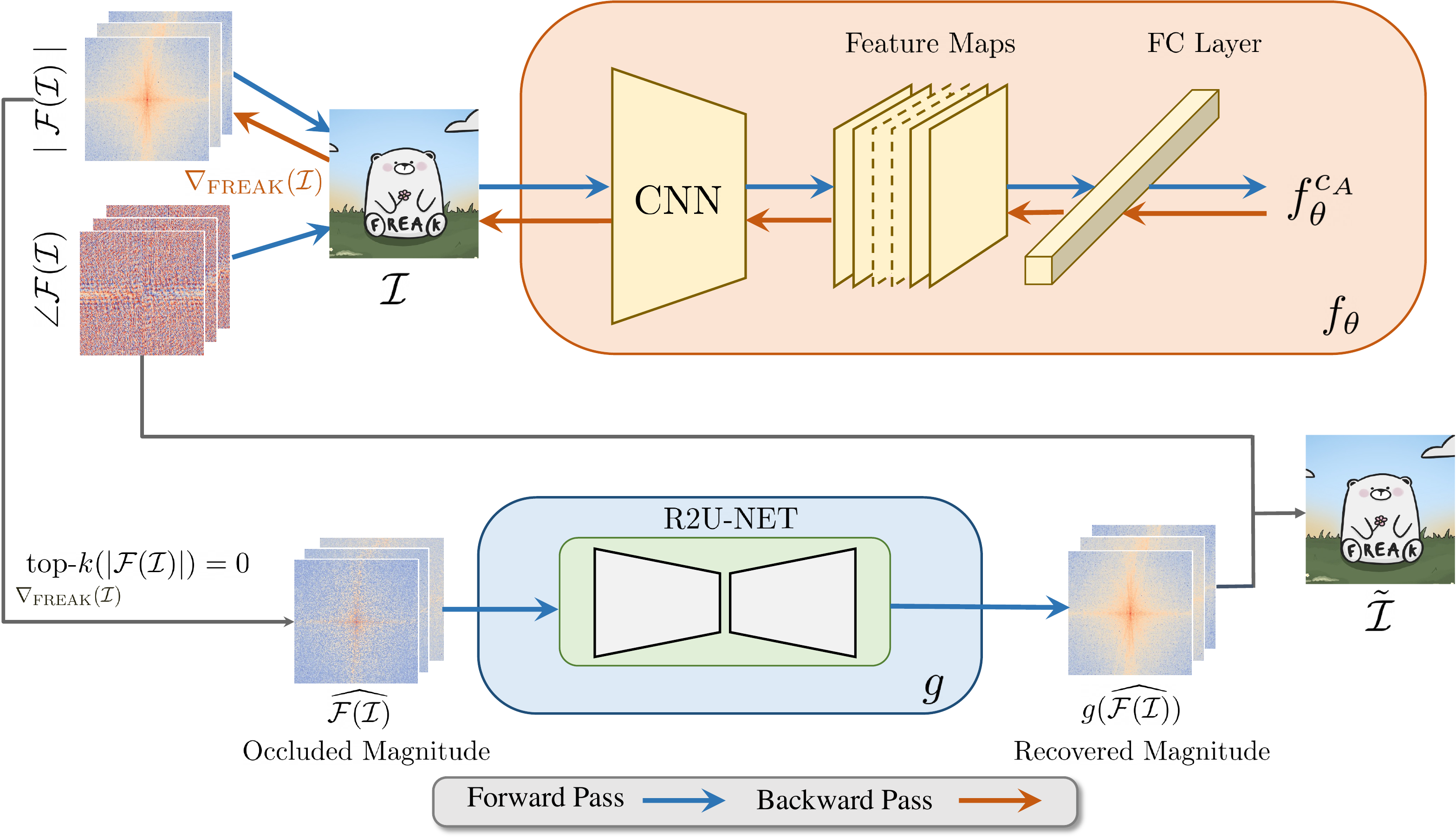}
    \caption{\textbf{FREAK for Image Purification.} We attempt to purify the images by first detecting the $k$ most sensitive frequency components, masking them by zero (occluding them), and then reconstructing the magnitude components using an R2U-Net.}
    \label{pipeline2}
\end{figure}

Given that $\hat{\nabla}_{\text{FREAK}}(x)$ allows us to locate the indices of top-$k$ most sensitive frequencies, a question that arises is can we reconstruct those magnitude values similar to what was done in Februus \cite{Doan2020FebruusIP}? Our results show that indeed this might be a feasible approach. This approach is summarized in Figure \ref{pipeline2}.

In simple terms, the idea is to locate the $k$ most sensitive frequency components, set them to zero and attempt to reconstruct them using an R2U-Net \cite{Alom2018RecurrentRC}. The loss used for training this network is a simple MSE on the recovered images and an MSE on the log Fourier recovered magnitude. Mathematically, the loss can be written as $$\mathcal{L} = \mathcal{L}_{\text{MSE}}(\Tilde{\mathcal{I}}, \mathcal{I}) + \mathcal{L}_{\text{MSE}}(\log(g(|\widehat{\mathcal{F}(\mathcal{I}})|)), \log(|{\mathcal{F}(\mathcal{I})}|)) $$

\begin{table*}[h!]

    \begin{subtable}[h]{0.32\textwidth}                \renewcommand{\arraystretch}{1.0} 
\scalebox{0.73}{
\begin{tabular}{@{}lcc@{}}
\toprule
\textbf{Method}  & \multicolumn{1}{c}{\textbf{CDA}\small{(\%)}} & \multicolumn{1}{c}{\textbf{ASR}\small{(\%)}} \\ \midrule
No Defense       & 92.51                            & 96.54                            \\
Gaussian (3x3)   & 65.12                            & 92.85                            \\
Gaussian (5x5)   & 36.07                            & 92.85                            \\
Weiner (3x3)     & 65.86                            & 92.85                            \\
Weiner (5x5)     & 42.58                            & 92.85                            \\
Highpass         & 31.34                            & 14.28                            \\
Lowpass          & 33.16                            & 71.42                            \\
Bandpass         & 22.98                            & 0.00                                \\
JPEG Compression & 83.80                             & 92.85                            \\
Autoencoder      & 82.33                            & 71.43                            \\ \midrule
\texttt {FREAK} (ours)   &                                  &                                  \\
\quad top-$k$ = 0\%         & 91.79                            & 85.71                            \\
\quad top-$k$ = 25\%        & 90.02                            & 14.55                            \\
\quad top-$k$ = 50\%        & 87.10                            & 5.59                                                     \\ \bottomrule
\end{tabular}}

        \caption{FIBA \cite{Feng2021FIBAFB}}
        \label{tab:week2}
     \end{subtable}
    \begin{subtable}[h]{0.32\textwidth}
        \centering
                \renewcommand{\arraystretch}{1.0} 
\scalebox{0.73}{
\begin{tabular}{@{}lcc@{}}
\toprule
\textbf{Method}  & \multicolumn{1}{c}{\textbf{CDA}\small{(\%)}} & \multicolumn{1}{c}{\textbf{ASR}\small{(\%)}} \\ \midrule
No Defense       & 92.84                            & 100.00                              \\
Gaussian (3x3)   & 64.00                               & 0.00                                \\
Gaussian (5x5)   & 34.80                             & 7.14                             \\
Weiner (3x3)     & 67.14                            & 7.14                             \\
Weiner (5x5)     & 46.93                            & 0.00                                \\
Highpass         & 33.81                            & 85.71                            \\
Lowpass          & 32.74                            & 7.14                             \\
Bandpass         & 26.72                            & 0.00                                \\
JPEG Compression & 85.05                            & 0.00                                \\
Autoencoder      & 82.05                            & 0.00                                \\ \midrule
\texttt {FREAK} (ours)   &                                  &                                  \\
\quad top-$k$ = 0         & 92.68                            & 100.00                           \\
\quad top-$k$ = 25\%        & 91.06                            & 0.85                            \\
\quad top-$k$ = 50\%        & 89.22                            & 0.65                                \\ \bottomrule
\end{tabular}}
       \caption{FTrojan \cite{Wang2021BackdoorAT}}
       \label{tab:week1}
    \end{subtable}
    \begin{subtable}[h]{0.32\textwidth}
        \centering
                \renewcommand{\arraystretch}{1.0} 
\scalebox{0.73}{
\begin{tabular}{@{}lcc@{}}
\toprule
\textbf{Method}  & \multicolumn{1}{c}{\textbf{CDA}\small{(\%)}} & \multicolumn{1}{c}{\textbf{ASR}\small{(\%)}} \\ \midrule
No Defense       & 94.43                            & 100.00                              \\
Gaussian (3x3)   & 63.96                            & 0.00                                \\
Gaussian (5x5)   & 29.38                            & 14.28                            \\
Weiner (3x3)     & 68.12                            & 7.14                             \\
Weiner (5x5)     & 47.84                            & 7.14                             \\
Highpass         & 36.96                            & 64.28                            \\
Lowpass          & 26.41                            & 7.14                             \\
Bandpass         & 27.14                            & 0.00                                \\
JPEG Compression & 86.91                            & 0.00                                \\
Autoencoder      & 83.15                            & 0.00                               \\ \midrule
\texttt {FREAK} (ours)   &                                  &                                  \\
\quad top-$k$ = 0         & 94.38                            & 92.89                            \\
\quad top-$k$ = 25\%        & 93.54                            & 7.03                            \\
\quad top-$k$ = 50\%        & 92.29                            & 5.59                                \\ \bottomrule
\end{tabular}}
       \caption{CYO \cite{Hammoud2021CheckYO}}
       \label{tab:week1}
    \end{subtable}
     \caption{\textbf{Defending Against Frequency Backdoor Attacks (CIFAR10).} The results of applying various defenses against existing frequency backdoor attacks show that using FREAK approach allows for the best balance between CDA and ASR.}
     \label{tab1v}
\end{table*}

We test that approach against a various set of defenses such as filtering approaches, some of which were proposed in  \cite{Hammoud2021CheckYO,Wang2021BackdoorAT}, namely,  Gaussian, Weiner, Highpass, Lowpass, and Bandpass filtering and compression approaches such as: JPEG and Autoencoder compression. As shown in Table \ref{tab1v}, this approach proves to be a solid approach to defend against backdoor attacks. More precisely, using this frequency reconstruction approach during test time allows us to maintain a high clean data accuracy while dropping the attack success rate to a very low level. This is observed for all three studied frequency-backdoor attacks.

However, our experiments on ImageNet show that this approach might not be scalable on large scale images where we observed a large drop in performance in terms of CDA and ASR trade-off. This calls for further research to develop a different loss function and architecture for applying frequency reconstruction.

\section{Additional Results:}

\subsection{Results on ResNet34}

In this subsection we present evaluations of FREAK against the CYO, FTrojan and FIBA using ResNet34 model instead of ResNet18. FREAK still proves to be a useful defense for detecting poisoned samples.

\begin{table}[h!]
\centering
\scalebox{0.9}{
\begin{tabular}{@{}ccc@{}}
\cmidrule(l){2-3}
\textbf{}        & \textbf{TPR (\%)} & \textbf{FPR (\%)} \\ \midrule
\textbf{CYO}     & 99.61             & 1.95              \\
\textbf{FTrojan} & 99.80            & 4.29              \\
\textbf{FIBA}    & 91.20             & 7.31              \\ \bottomrule
\end{tabular}}
\caption{\textbf{FREAK Against Frequency Backdoor Attacks.} FREAK proves to be capable of achieving a high TPR while maintaining a low FPR against frequency backdoor-attacks.}
\end{table}

\subsection{Hyperparameter Sensitivity of FREAK}

Now we study the sensitivity of FREAK to the different hyperparameters. Unless the hyperparameter is being ablated, the value is fixed to that presented in the manuscript \ie $\alpha=1$, $|\mathcal{D}_h|= 32$, $|\mathcal{D}_c| = 128$, $\beta = 12$, and $k=5000$.

\paragraph{Top-$k$ Value.}

As shown in table \ref{topk}, increasing the value of $k$ allows for a lower FPR at the cost of a drop in TPR.

\begin{table}[h!]
\centering
\begin{tabular}{@{}cccc@{}}
\cmidrule(l){2-4}
                                  & \textbf{$k$} & \textbf{TPR (\%)} & \textbf{FPR (\%)} \\ \midrule
\multirow{2}{*}{\textbf{CYO}}     & 2500             & 99.68             & 2.97              \\
                                  & 7500             & 99.02             & 1.95              \\ \midrule
\multirow{2}{*}{\textbf{FTrojan}} & 2500             & 100.00            & 3.90              \\
                                  & 7500             & 100.00            & 2.73              \\ \midrule
\multirow{2}{*}{\textbf{FIBA}}    & 2500             & 87.89             & 5.85              \\
                                  & 7500             & 85.46             & 5.63              \\ \bottomrule
\end{tabular}
\caption{\textbf{Effect of $k$ in top-$k$ Operation of FREAK}}
\label{topk}
\end{table}

\clearpage
\paragraph{Size of Pooling Filter ($\beta$)}

Table \ref{beta}, shows the effect of changing the filter size $\beta$.

\begin{table}[h!]
\centering
\begin{tabular}{@{}cccc@{}}
\cmidrule(l){2-4}
                                  & \textbf{$\beta$} & \textbf{TPR (\%)} & \textbf{FPR (\%)} \\ \midrule
\multirow{2}{*}{\textbf{CYO}}     & 9                & 99.22             & 1.57              \\
                                  & 16               & 99.22             & 1.95              \\ \midrule
\multirow{2}{*}{\textbf{FTrojan}} & 9                & 100.00            & 2.34              \\
                                  & 16               & 100.00            & 4.29              \\ \midrule
\multirow{2}{*}{\textbf{FIBA}}    & 9                & 91.79             & 8.98              \\
                                  & 16               & 88.08             & 4.98              \\ \bottomrule
\end{tabular}
\caption{\textbf{Effect of Pooling Size $\beta$ on FREAK}}
\label{beta}
\end{table}

\paragraph{Size of Held Out Set}

Table \ref{heldout}, shows the effect of increasing the size of the held-out set. Our results show little to no change in the performance of FREAK with increased size of held-out set.

\begin{table}[h!]
\centering
\begin{tabular}{@{}cccc@{}}
\cmidrule(l){2-4}
                                  & \textbf{$\mid \mathcal{D}_h \mid$} & \textbf{TPR (\%)} & \textbf{FPR (\%)} \\ \midrule
\multirow{2}{*}{\textbf{CYO}}     & 64                                 & 99.22             & 1.95              \\
                                  & 256                                & 99.22             & 1.95              \\ \midrule
\multirow{2}{*}{\textbf{FTrojan}} & 64                                 & 100.00            & 2.73              \\
                                  & 256                                & 100.00            & 2.73              \\ \midrule
\multirow{2}{*}{\textbf{FIBA}}    & 64                                 & 89.68             & 5.15              \\
                                  & 256                                & 90.47             & 5.31              \\ \bottomrule
\end{tabular}
\caption{\textbf{Effect of Size of Held-Out Set on  FREAK}}
\label{heldout}
\end{table}

\paragraph{Size of Clean Set}

Table \ref{clean}, shows the effect of changing the size of the clean set. Our results show little to no change in the performance of FREAK with increased size of held-out set.

\begin{table}[h!]
\centering
\begin{tabular}{@{}cccc@{}}
\cmidrule(l){2-4}
                                  & \textbf{$\mid \mathcal{D}_c \mid$} & \textbf{TPR (\%)} & \textbf{FPR (\%)} \\ \midrule
\multirow{2}{*}{\textbf{CYO}}     & 64                                 & 99.22             & 2.23              \\
                                  & 256                                & 99.22             & 1.56              \\ \midrule
\multirow{2}{*}{\textbf{FTrojan}} & 64                                 & 100.00            & 2.45              \\
                                  & 256                                & 100.00            & 2.73              \\ \midrule
\multirow{2}{*}{\textbf{FIBA}}    & 64                                 & 86.48             & 4.02              \\
                                  & 256                                & 90.53             & 5.47              \\ \bottomrule
\end{tabular}
\caption{\textbf{Effect of Size of Clean Set on  FREAK}}
\label{clean}
\end{table}

\clearpage

\paragraph{Trade-Off Parameter $\alpha$}

Table \ref{alpha}, shows the effect of changing the confidence parameter $\alpha$. As expected, as $\alpha$ increases we are less

\begin{table}[h!]
\centering
\begin{tabular}{@{}cccc@{}}
\cmidrule(l){2-4}
                                  & \textbf{$\alpha$} & \textbf{TPR (\%)} & \textbf{FPR (\%)} \\ \midrule
\multirow{2}{*}{\textbf{CYO}}     & 2                              & 99.02             & 0.78              \\
                                  & 4                              & 96.09             & 0.39              \\ \midrule
\multirow{2}{*}{\textbf{FTrojan}} & 2                              & 100.00            & 1.17              \\
                                  & 4                              & 100.00            & 1.17              \\ \midrule
\multirow{2}{*}{\textbf{FIBA}}    & 2                              & 81.49             & 2.96              \\
                                  & 4                              & 65.00             & 1.25              \\ \bottomrule
\end{tabular}
\caption{\textbf{Effect of Changing $\alpha$ on FREAK}}
\label{alpha}
\end{table}

\end{document}